\documentclass[prc,onecolumn,nofootinbib,showpacs,floatfix,a4paper]{revtex4-1}

\usepackage{graphicx}
\usepackage{dcolumn}
\usepackage{bm}
\usepackage{amsmath,amssymb}
\usepackage[body={17.5cm,25cm},top=1.75cm]{geometry}
\usepackage{ulem,color}

\newcommand{\eq}[1]{\begin{equation} #1 \end{equation}}
\newcommand{\bra}[1]{\langle #1|}
\newcommand{\ket}[1]{|#1\rangle}
\newcommand{\elmx}[3]{\langle #1|#2|#3 \rangle}
\newcommand{\overlap}[2]{\langle #1 | #2 \rangle}

\newcommand{\Vres}{\hat V_{\rm res}}
\newcommand{\Hiqp}{\hat H_{\rm IQP}}

\long\def\symbolfootnotetext[#1]#2{\begingroup%
\def\thefootnote{\fnsymbol{footnote}}\footnotetext[#1]{#2}\endgroup}

\begin{document}


\title{
Parity restoration in the Highly Truncated Diagonalization
  Approach: application to the outer fission barrier of $^{240}$Pu}
\author{T. V. Nhan Hao$^{1, 2, 3,*}$, P. Quentin$^{2, 3}$ and
  L. Bonneau$^{2, 3}$}

\affiliation{
\textsuperscript{1} Tan Tao University, Tan Tao
  University Avenue, Tan Duc Ecity, Long An Province, Vietnam\\
\textsuperscript{2} Univ.\,Bordeaux, CENBG, UMR5797, F-33170, Gradignan,
France\\
\textsuperscript{3} CNRS, IN2P3, CENBG, UMR5797, F-33170, Gradignan,
France}

\date{\today}

\begin{abstract}

The restoration of the parity symmetry has
been performed in the framework of the Highly Truncated
Diagonalization Approach suited to treat correlations in an explicitly
particle-number conserving microscopic approach. To do so we have assumed axial symmetry and used a generalized Wick's theorem due to
L\"owdin in a projection-after-variation scheme. We have chosen
the Skyrme SkM$^*$ energy-density
  functional for the particle-hole channel and a density-independent
delta force for the residual interaction. We have applied this approach in the region of the
outer fission barrier of the $^{240}$Pu nucleus. As a result, we have shown that the $K^{\pi} = 0^+$ fission
isomeric state is statically unstable against intrinsic-parity
breaking modes, while the projection does not affect
the energy at the top of the intrinsic outer
fission barrier. Altogether, this leads to an increase of the height
of the outer fission barrier--with respect to
the fission isomeric state--by about 350 keV, affecting thus
significantly the fission-decay lifetime of the
considered fission isomer.

\end{abstract}


\maketitle


\section{Introduction}
\label{Intro}

\symbolfootnotetext[1]{Corresponding author: hao.tran@ttu.edu.vn}

It has been recognized very long ago that the second fission barriers
of heavy nuclei are asymmetric under left-right reflection (or
intrinsic parity) symmetry~\cite{PMSGN}. Early
self-consistent calculations \cite{Berger} allowing to violate this
symmetry have demonstrated that the height of the second fission
barrier of actinide nuclei could be substantially reduced in that
way. This has been confirmed in many further calculations (see, e.g.,
Ref.~\cite{BQS} using the Hartree--Fock+BCS approach). This phenomenon
is driven by the heavy fragment shell effects explaining
thus~\cite{PMSGN}, already at this early stage of the fission process,
the asymmetric pattern of the fragment mass yields (at low
compound-nucleus energy) which had been observed in this region long
before.

In most microscopic or macro-microscopic calculations in the Uranium,
Plutonium region, upon increasing the elongation after the fission
isomeric state, the intrinsic equilibrium solution becomes unstable
with respect to the left-right reflection symmetry and acquires
rapidly a larger and larger octupole deformation which is stabilized
at a value corresponding, as just noted, to the most probable
fragmentation. Yet, even though the intrinsic parity may be broken
within some auxiliary microscopic solutions under consideration, the
parity of the physical solution must be conserved during the fission
process. For instance, if one describes the spontaneous fission of an
even-even fissioning nucleus as $^{240}$Pu, one should evaluate the
fission barrier obtained upon projecting intrinsic solutions on the
desired parity.

The Highly Truncated Diagonalization Approach (HTDA) is designed to
produce realistic correlated wave functions in an approach which
conserves explicitly the particle number and does not violate the
Pauli principle \cite{Pillet2002}. While our approach takes stock of
state-of-the-art energy-density
functionals to describe microscopically mean-field properties, it had
been initiated long before in simpler terms by various groups (see,
e.g., Refs.~\cite{Zeng1983,Hata1987,Buro1996}). The aim of this paper
is to present the first application of this HTDA method in a
projected-after-variation calculation of the second fission barrier of
$^{240}$Pu where the octupole deformation mode plays a very important
role. It should be noted that very few microscopic calculations of fission barriers with parity-symmetry restoration have been
  performed so far. Among these calculations one can emphasize the
  systematic study of fission barriers carried out by Samyn and
  collaborators~\cite{Samyn05} within the
  Skyrme--Hartree--Fock--Bogolyubov approach with separate
  particle-number and parity projections using the BSk8
  parametrization.

This paper is organized as follows. In Sec. II, a detailed
presentation of the HTDA formalism and some of its key aspects are
given. The analysis of the numerical results is presented in
Sec. III. In the last section, the main results are summarized and
some conclusions are drawn.

%
%

\section{Theoretical framework}

%
%

\subsection{The Highly Truncated Diagonalization Approach}

First let us briefly recall some general features of the HTDA approach
which allows to treat on the same footing various correlations,
including pairing correlations, in an explicitly particle number
conserving approach as discussed in Refs. \cite{Pillet2002, Quentin2007, Houda2010,Lafchiev2010,Bonneau2007PRC}. In the next section the modifications to this standard approach, due to the projection method introduced in this paper,
will be discussed in detail.

We start from the Hamiltonian
\begin{equation}
\hat{H}=\hat{K}+\hat{V}
\end{equation}
where $\hat{K}$ is the kinetic energy and $\hat{V}$ a two-body
interaction. The latter may be density dependent as it is the case for
the Gogny or the Skyrme interactions and it includes the proton-proton
Coulomb interaction. In so far as the exchange term of the latter
interaction is treated within the Slater approximation, as it is
generally performed in practice, some further spurious density
dependence is included. For the sake of simplicity we will skip first
the density dependence of $\hat{V}$, the effect of which in practice
will be discussed later.

We introduce an a priori arbitrary one-body potential $\hat{U}$. Of
course, we will try and implement all what is possible to incorporate
in terms of guessed one-body properties of the correlated wave function
in this auxiliary mean field (generally through some simplified
self-consistent approach). We then introduce the corresponding
auxiliary one-body Hamiltonian
\begin{equation}
\hat{H}_{\mathrm{0}}=\hat{K}+\hat{U}
\end{equation}
and its lowest-energy eigenstate $\ket{\Phi_0}$ which is a Slater
determinant corresponding to a quasi-particle vacuum of particle-hole
type. In the following we will consider particle-hole excitations on
this vacuum and define normal products in terms of the related
quasi-particle operators.

Now we rewrite the original microscopic Hamiltonian as
\begin{equation}
\hat{H} = \Hiqp + \Vres +
\langle\mathrm{\Phi_0}|\hat{H}|\Phi_0\rangle \:,
\label{eq:hamilappro}
\end{equation}
where the independent quasi-particle Hamiltonian
$\hat{H}_{\mathrm{IQP}}$ and the residual interaction $\Vres$ are
defined by
\begin{align}
\Hiqp & = \hat{H}_{0}-\langle\mathrm{\Phi_0}|\hat{H}_0|\Phi_0\rangle
\\
\Vres & = (\hat{V}-\hat{U})-\elmx{\Phi_0}{(\hat{V}-\hat{U})}{\Phi_0} \:.
\end{align}
The Wick theorem applied to one- and two-body operators such as
$\hat{U}$ and $\hat{V}$, respectively, implies that
\begin{align}
\hat{U} & = \, :\hat{U}: + \, \elmx{\Phi_0}{\hat{U}}{\Phi_0} \\
\hat{V} & = \, :\hat{V}: + :\overline{V}: + \, \elmx{\Phi_0}{\hat
  V}{\Phi_0} \:,
\end{align}
where $\overline{V}$ denotes the one-body reduction of $\hat{V}$ for
the particle-hole vacuum $\ket{\Phi_0}$. Consequently one has
\begin{align}
\Hiqp & = \, :\hat H_0: \\
\Vres & = \, :\hat{V}: \, + \, :\overline{V} - \hat{U}: \:.
\label{eq:vreshtda}
\end{align}
Assuming that $\hat{V}$ and $\hat{U}$ reasonably produce similar
descriptions of one-body properties of the studied nuclear system,
we make the approximation of neglecting the $:\overline{V} - \hat{U}:$
term.

In what follows we will use for $\hat{V}$ a Skyrme nucleon-nucleon
effective interaction (adding of course the Coulomb interaction). As
well known however, most of the parametrizations of
such forces are not suited to treat correctly pairing correlations,
with some noticeable exceptions though, as, e.g., in
the parametrization SkP of Ref.~\cite{Dobaczewski84} and SGII
of Ref.~\cite{Giai1981}. In order to correct this deficiency one
replaces the two-body interaction $\hat V$ appearing in the definition
(\ref{eq:vreshtda}) of $\Vres$ by a zero-range interaction
$\hat{V}_{\delta}$ as it is usual for the description of such
nuclear correlations. We choose it density-independent since the ad
hoc introduction here of a specific surface dependence is neither
theoretically ascertained nor phenomenologically
justified. Specifically, while the Skyrme force will be used in the
particle-hole channel (in $\hat{H}_{\mathrm{IQP}}$ and to evaluate
$\langle\mathrm{\Phi_0}|\hat{H}|\Phi_0\rangle$), the $\delta$ force
will be used for the particle-particle (and hole-hole) channel. In
practice one thus approximates $\Vres$ by
\begin{equation}
\Vres \approx \hat{V}_{\delta} - :\overline{V}_{\delta}: -
\elmx{\Phi_0}{\hat{V}_{\delta}}{\Phi_0} \:,
\label{eq:deltaprro}
\end{equation}
where $\overline{V}_{\delta}$ is the one-body reduction of the
$\hat{V}_{\delta}$ interaction for the particle-hole vacuum
$\ket{\Phi_0}$.

It is to be noticed that the above replacement of the two-body
interaction introduces some state dependence of the
Hamiltonian. In fact, introducing such an unwanted feature is
unfortunately common, already at the Hartree-Fock level, in the
current state-of-the-art microscopic treatments of nuclear properties
when using Gogny or Skyrme energy-density functionals. The relevance
of such a seemingly dubious approach should be ultimately validated,
as usual, by the phenomenological quality of the results so obtained.

The Hamiltonian (\ref{eq:hamilappro}) is diagonalized in the
orthonormal many-body basis which includes the particle-hole vacuum
$\ket{\Phi_0}$ and $n$-particle-$n$-hole ($n$p$n$h) excitations with
respect to $\ket{\Phi_0}$, generically noted $\ket{\Phi_n}$, with $n$
ranging from 1 to some maximum order $m$. The correlated ground state
can thus be written as
\begin{equation}
\ket{\Psi} = \chi_0 \, \ket{\Phi_0} + \sum_{n=1}^m
\sum_{i} \chi_n^{(i)} \, \ket{\Phi_n^{(i)}} \:,
\end{equation}
where $i$ distinguishes the various $n$p$n$h configurations.

A satisfactory HTDA description of pairing correlations in heavy
stable nuclei has been generally obtained with a many-body
basis involving the particle-hole vacuum $\ket{\Phi_0}$ and pair
excitations only. In this context the Cooper pairs imply strictly
time-reversed states, so that they have in particular the same
charge. This approach deals then with the so-called $|T_z|=1$ or $nn$,
$pp$ pairing correlations, an approach which is relevant far from the
$N\!=\!Z$ line. In contrast the $np$ pairing correlations are
important when studying nuclei near the $N\!=\!Z$ line (see, e.g., a
recent thorough study in Ref.~\cite{Julien2012} within the HTDA
framework).

Given the two approximations which have been made in
$\Vres$ (regarding $:\!\overline{V}-\hat{U}\!:$ and
$:\!\hat{V}\!:$), the effect of including density-dependent parts in the
2-body interaction is limited to the evaluation of $\langle
\mathrm{\Phi_0}|\hat{H}|\mathrm{\Phi_0}\rangle$ where they have been
fully taken into account.

%
%

\subsection{Projected energy}
\label{proenergy}

The state $\ket{\Psi_p}$ of definite parity $p=\pm 1$ is obtained by
the action of the parity projection operator $\hat{P}_p =
\frac{1}{2}(1 + p \,\hat{\mathrm{\Pi}})$, where $\hat{\Pi}$ denotes
the parity operator, onto the correlated (left-right
asymmetrical) state $\ket{\Psi}$
\begin{equation}
\ket{\Psi_p} = N_p \hat{P}_p \ket{\Psi} \:,
\end{equation}
where $N_p$ is a real normalization constant defined by
\begin{equation}
N_p = \sqrt 2 \, \big(1+p \, \elmx{\Psi}{\hat{\Pi}}{\Psi}\big)^{-\frac
  1 2} \:.
\end{equation}
The resulting projected energy $E^{(p)} = \elmx{\Psi_p}{\hat H}{\Psi_p}$
thus takes the form
\eq{
E^{(p)} = \frac{\elmx{\Psi}{\hat H}{\Psi} + \elmx{\widetilde{\Psi}}{\hat
    H}{\widetilde{\Psi}} + p \, \big(\elmx{\Psi}{\hat
    H}{\widetilde{\Psi}} + \elmx{\widetilde{\Psi}}{\hat H}{\Psi}\big)
  }{2\big(1 + \overlap{\Psi}{\widetilde{\Psi}}\big)} \:,
\label{eq:energytotal}
}
with $\ket{\widetilde{\Psi}} = \hat{\Pi} \ket{\Psi}$.

The Hamiltonian $\hat{H}$ including a density-dependent effective
interaction breaks the parity symmetry as long as the density does
it. So it seems contradictory to restore a symmetry for eigensolutions
of a Hamiltonian which breaks also this specific symmetry. Such a
feature is of course not related nor restricted to the HTDA approach,
but is a consequence of the use of solution-dependent Hamiltonians as
we have seen above when using, e.g., Skyrme or Gogny
energy-density functionals. Moreover the
calculation of $E^{(p)}$ is rather involved since it requires to
calculate four matrix elements together with one overlap. For these
two reasons one has chosen here
to make some simplifying assumptions on the hamiltonian and
overlap kernels explained in the next subsection.

%
%

\subsection{Approximate Hamiltonian and overlap kernels}
\label{simplemodel}

In principle, there is a total arbitrariness in choosing the one-body
potential $\hat{U}$ to be introduced in the HTDA Hamiltonian. However
our solution of the secular equation is expanded in a truncated basis
spanned by a subset of $A$-particle Slater determinants. On the one
hand such a limitation is embodied by the so-called single-particle
configuration space, which is the restricted set of eigenfunctions of
the one-body Hamiltonian $\hat{H}_0$ used to define particle and hole
states. On the other hand the many-body configuration space is
restricted, beyond the quasi-particle vacuum $\ket{\Phi_0}$, to
one-pair excitations with respect to $\ket{\Phi_0}$ written with a
transparent notation as
\begin{equation}
\ket{\Phi_2}= a^{\dagger}_{k'}a^{\dagger}_{\bar{k}'} a_{\bar{k}}a_k
\ket{\Phi_0} \:.
\label{eq:phi2}
\end{equation}
As a result our effective Hamiltonian does depend on the choice of the
one-body potential $\hat{U}$.

Another $\hat{U}$-dependence of the Hamiltonian stems from the replacement of the Skyrme two-body interaction
by a delta force in the definition of the residual interaction
$\Vres$. The one-body reduction of $\hat{V} - \hat{V}_{\delta}$ with
respect to the particle-hole vacuum state $\ket{\Phi_0}$
depends by definition on this state and
then ultimately on the potential $\hat{U}$. Similarly this replacement
induces a $\hat{U}$-dependence of the expectation value of
$\hat{V}-\hat{V}_{\delta}$ for $\ket{\Phi_0}$.

If the Hamiltonian were truly independent of the
single-particle potential $\hat U$, we would have the freedom of
choosing $\hat{U}$ according to the type of matrix elements involved
in the calculation of $E^{(p)}$. Indeed, $\hat U$ does not have to assume
a unique functional form and can be defined in a piecewise
way in the one-body configuration space. Because of the above
discussed limitations, doing so becomes an approximation. We explain
below what choice we make to evaluate the Hamiltonian and and overlap
kernels.

\subsubsection{Diagonal terms of the hamiltonian kernel}

The matrix element $\elmx{\Psi}{\hat H}{\Psi}$ is encountered
in usual HTDA calculations. To
  calculate the contribution from $\Hiqp$, we choose the one-body potential $\hat U$ as the one-body
reduction of $\hat{V}$ associated with the particle-hole vacuum $\ket{\Phi_0}$.

In the case of the matrix element $\elmx{\widetilde{\Psi}}{
    \hat H}{\widetilde{\Psi}}$, we replace the above potential
$\hat{U}$ with the parity-transformed operator $\widetilde{U}$ defined
  by
\eq{
\widetilde{U} = \hat{\mathrm{\Pi}} \hat{U} \hat{\mathrm{\Pi}} \:.
}
The corresponding particle-hole vacuum is thus nothing but the mirror
image $\ket{\widetilde{\Phi}_0} = \hat{\Pi} \ket{\Phi_0}$ of the
particle-hole vacuum $\ket{\Phi_0}$ involved in the calculation of
$\elmx{\Psi}{\hat H}{\Psi}$. Then the one-body reduction of
$\hat{V}_{\delta}$ for $\ket{\widetilde{\Phi}_0}$ is the
parity-transformed operator $\widetilde{\overline{V}}_{\delta} =
\hat{\mathrm{\Pi}} \overline{V}_{\delta} \hat{\mathrm{\Pi}}$. The
resulting $\Vres$ operator takes the form $\Vres = \hat{V}_{\delta} -
:\widetilde{\overline{V}}_{\delta}: -
\elmx{\widetilde{\Phi}_0}{\hat{V}_{\delta}}{\widetilde{\Phi}_0}$, and the
corresponding HTDA Hamiltonian $\widetilde{H}$ is thus simply
related to the previous Hamiltonian by a parity transformation. Owing
to the unitary character of $\hat{\Pi}$ one thus obtains
\begin{equation}
\elmx{\widetilde{\Psi}}{\widetilde{H}}{\widetilde{\Psi}} =
\elmx{\Psi}{\hat H}{\Psi} \:.
\label{eq:psihpsi}
\end{equation}

\subsubsection{Off-diagonal terms of the hamiltonian kernel}

The off-diagonal matrix elements $\elmx{\Psi}{\hat
  H}{\widetilde{\Psi}}$ and $\elmx{\widetilde{\Psi}}{\hat H}{\Psi}$
are numerically difficult to evaluate,
especially in the context of the above discussed state-dependence of
the Hamiltonian.

First of all, by making a judicious
choice of the arbitrary one-body potential $\hat{U}$ entering the
definition of the HTDA Hamiltonian, it is possible in general to
equate the two matrix elements $\elmx{\Psi}{\hat
  H}{\widetilde{\Psi}}$ and $\elmx{\widetilde{\Psi}}{
  {\widetilde{H}}}{\Psi}$. For that it suffices to
choose $\widetilde{U}$ and the resulting
  particle-hole vacuum $\ket{\widetilde{\Phi}_0}$ similarly to what has been
discussed for the diagonal term $\elmx{\widetilde{\Psi}}{\hat
    {H}}{\widetilde{\Psi}}$. Taking advantage of the arbitrariness of
the auxiliary one-body potential used to define the HTDA Hamiltonian, one may decide to
consider the Hamiltonian $\widetilde  {H}$ for matrix elements between
the bra $\bra{\Psi}$ and the ket $\ket{\widetilde{\Psi}}$, while using
the Hamiltonian $\hat  {H}$ for matrix elements between the bra
$\bra{\widetilde{\Psi}}$ and the ket $\ket{\Psi}$.

Actually, we have performed a further approximation prompted by two
motivations. First, as already noted, we introduce in practice through
our approximate diagonalization process a spurious dependence of  the
Hamiltonian of the auxiliary one-body potential $\hat {U}$. This makes
the above choice somewhat arbitrary. Second,
we face here the problem usually encountered in
configuration-mixing calculations using state-dependent
Hamiltonians. It concerns, as well known, the choice of the reference
state defining this Hamiltonian when computing non-diagonal matrix
elements (where non-diagonal refers here to different reference
states). One could think of using the one-body reduction of the
underlying two-body interaction with respect to the mixed density in
the sense of, e.g., Ref.~\cite{GCM}. But, in practice, it so happens
that the non-negative character of the corresponding local
density (its matrix elements which are diagonal in $\mathbf{r}$-space)
is not always warranted. This entails, of
course, a problem to define the density-dependent part of the Skyrme
mean field and the Slater Coulomb exchange mean field. These fields
involve, often for the former and surely for the
latter, a non-integer power of the density. Moreover let us state
again that it does not seem fully satisfactory, a priori, to use an
auxiliary Hamiltonian breaking some symmetry in the process of
restoring this symmetry for its solution.

This is why we have made a
further use of the arbitrariness of the one-body operator $\hat{U}$
as follows. For a given intrinsic solution in the collective
deformation space as specified, e.g., by its axial quadrupole $Q_{20}$
and octupole $Q_{30}$ moments, we define the potential $\hat U$ in the
HTDA Hamiltonian as the one corresponding to the parity-symmetrical
vacuum solution ($Q_{30}=0$) with the
same value of the axial quadrupole moment.

In that case, since the two Hamiltonians $\hat {H}$ and $\widetilde
{H}$ are trivially identical, one gets
\begin{equation}
\langle \mathrm{\Psi}|\hat{H}|\mathrm{\widetilde{\Psi}}\rangle =
\langle \mathrm{\widetilde{\Psi}}|\hat{H}|\mathrm{\Psi}\rangle \:,
\label{eq:psihpipsi}
\end{equation}
and since these matrix elements are real, we deduce that the
HTDA Hamiltonian is indeed hermitian.

Using Eqs.~(\ref{eq:psihpsi}) and (\ref{eq:psihpipsi}) we can
rewrite the projected energy (\ref{eq:energytotal}) as
\begin{equation}
\label{Ep}
E^{(p)}=\frac{\langle
  \mathrm{\Psi}|\hat{H}|\mathrm{\Psi}\rangle+p\langle
  \mathrm{\Psi}|\hat{H}|\widetilde{\mathrm{\Psi}}\rangle}{1+p \,
  \langle \mathrm{\Psi}|\widetilde{\mathrm{\Psi}}\rangle} \:.
\end{equation}

To compute the matrix elements involving bras $\langle\mathrm{\Psi}|$
and kets $|\widetilde{\mathrm{\Psi}}\rangle$, we take stock of the fact that we have to deal with matrix elements between Slater
  determinants. It is thus more appropriate to use the L\"owdin approach of
  Ref.~\cite{Lowdin1} than the Balian-Brezin generalized Wick theorem
  \cite{BB}. Let us denote by $\ket{\Phi}$ and $\ket{\Phi'}$
  two Slater determinants built from single-particle states
  generically noted $\ket{\phi_k}$ and $\ket{\phi'_l}$,
  respectively, and by $\mathbf d$ the corresponding overlap matrix. By
definition the elements $d_{kl}$ of the overlap matrix are the
overlaps $\overlap{\phi_k}{\phi'_l}$ of the occupied states
$\ket{\phi_k}$ of the Slater determinant $\ket{\Phi}$ with the
occupied states $\ket{\phi'_l}$ of $\ket{\Phi'}$. Then the matrix
elements of a one-body operator $\hat V_1$ and a two-body operator
$\hat V_2$ between $\ket{\Phi}$ and $\ket{\Phi'}$ are given
by~\cite{Lowdin1}
\begin{align}
\elmx{\Phi}{\hat V_1}{\Phi'} = &
\sum_{\substack{k \in \Phi \\ l \in \Phi'}} (-1)^{k + l}
\elmx{\phi_k}{\hat V_1}{\phi'_l} \, D_{\Phi \Phi'}(k|l) \\
\elmx{\Phi}{\hat V_2}{\Phi'} = & \frac{1}{2}
\sum_{\substack{k_1 \ne k_2 \in \Phi \\ l_1 \ne l_2 \in \Phi'}}
(-1)^{k_1+k_2 + l_1+l_2} \elmx{\phi_{k_1}\phi_{k_2}}{\hat
    V_1}{\phi'_{l_1}\phi'_{l_2}} \times \nonumber \\
& \phantom{\sum_{\substack{k_1 \ne k_2 \in \Phi \\ l_1 \ne l_2 \in \Phi'}}}
 D_{\Phi \Phi'}(k_1k_2|l_1l_2) \:,
\end{align}
where the sums run over the occupied states of the Slater
determinants and $D_{\Phi \Phi'}(k_1\cdots k_n|l_1\cdots l_n)$ is the
minor of order $n$ of the overlap matrix associated with the
lines $k_1$, ..., $k_n$ and columns $l_1$, ..., $l_n$ (i.e. the
determinant of the submatrix obtained by removing the lines $k_1$,
..., $k_n$ and columns $l_1$, ..., $l_n$ from the overlap matrix).

\subsubsection{Overlap kernel}

Within the HTDA framework the overlap
$\overlap{\Psi}{\widetilde{\Psi}}$ can be decomposed as
\begin{align}
\langle\Psi |\widetilde{\Psi}\rangle = &
\chi_{0}^2 \langle \Phi_{0}|\widetilde{\Phi}_{0}\rangle+2\chi_{0}
\sum_{k=1}^{{N-1}}\chi^{(k)}_{2}
\langle\Phi_{0}|\widetilde{\Phi}^{(k)}_{2}\rangle \nonumber \\
 & + 2\sum_{k=1}^{N-1}\sum_{l=1}^{k-1}\chi^{(k)}_{2}
\chi^{(l)}_{2} \langle\Phi^{(k)}_{2}|\widetilde{\Phi}^{(l)}_{2}\rangle
\nonumber\\
& +\sum_{k=1}^{N-1}(\chi^{(k)}_2)^2 \langle
\Phi^{(k)}_{2}|\widetilde{\Phi}^{(k)}_{2}\rangle  \:.
\label{eq:recouhtda}
\end{align}
The sums run over the $N-1$ one-pair configurations, where $N$
is the size of the many-body basis.

As demonstrated by L\"owdin \cite{Lowdin1} the overlap of any two
Slater determinants $\ket{\Phi}$ and $\ket{\Phi'}$ can be expressed as
the determinant of their overlap matrix $\mathbf{d}$:
\begin{equation}
\overlap{\Phi}{\Phi'} = \det \mathbf{d} \:.
\end{equation}

The time-reversal invariance of the $\Hiqp$
Hamiltonian allows us to split the space of
one-body states in two subspaces $\mathcal E$ and $\mathcal E^*$
deduced one from the other by time-reversal symmetry. Due to the axial
symmetry we may define $\mathcal E$ as the
subspace which incorporates all single-particle states $\ket{\phi_i}$
such that $\elmx{\phi_i}{\hat j_z}{\phi_i} >0$. The matrix
$\mathbf{d}$ can be written as a block-diagonal matrix defined
as
\begin{equation}
\mathbf{d}=
\begin{pmatrix}
 D & 0\\
  0  & D^*
 \end{pmatrix}
\end{equation}
where $D^*$ is the complex conjugate of the submatrix $D$, so
that $\overlap{\Phi}{\Phi'} = \big|\det D\big|^2$.
Moreover, since all
the occupied states involved in $D$ are
  eigenstates of the $\hat j_z$ operator with positive eigenvalues
$\Omega_m$ (in $\hbar$ unit), the matrix $D$ is block diagonal:
\eq{
D = \begin{pmatrix}
D_{\Omega_1} & 0 & \cdots & & \\
0 & D_{\Omega_2} & 0 & \cdots & \\
\vdots & & \ddots & & \vdots \\
 & & 0 & D_{\Omega_m} & 0 \\
 & & & \cdots & 0
\end{pmatrix} \:,
}
where the zeros have to be understood as null submatrices and
  the submatrices $D_{\Omega_1}$, ...,
$D_{\Omega_m}$ correspond to $\Omega$-values for which the
  Slater determinants $\ket{\Phi}$ and $\ket{\Phi'}$ involve a finite
  number of occupied states with the quantum number $\Omega$. If
  $s_{\Omega}$ ($s'_{\Omega}$ resp.) denotes the number of occupied
  states in $\ket{\Phi}$ ($\ket{\Phi'}$ resp.) with the eigenvalue
  $\Omega$ of the $\hat j_z$ operator, then the dimension of the
  submatrix $D_{\Omega}$ is the minimum of $s_{\Omega}$ and
  $s'_{\Omega}$. If $s_{\Omega} \ne s'_{\Omega}$ the overlaps
  involving the occupied states of $\ket{\Phi}$ or $\ket{\Phi'}$--with
  the eigenvalue $\Omega$--which do not appear in the
  submatrix $D_{\Omega}$ vanish and so does the determinant of
  $D$. Therefore the non-vanishing overlaps $\overlap{\Phi}{\Phi'}$
  are those for which $s_{\Omega} = s'_{\Omega}$ for all
  $\Omega$-values occuring in both $\ket{\Phi}$ and $\ket{\Phi'}$. In
  this case, the overlap may be written as
\begin{equation}
\overlap{\Phi}{\Phi'} = \prod_m \big|\det D_{\Omega_m}\big|^2 \:,
\end{equation}
where the product runs over all positive values of $\Omega$
involved in $\ket{\Phi}$ and $\ket{\Phi'}$.

\subsection{Some numerical aspects}

The computational process which has been followed may be decomposed in
three steps.

First we perform usual Hartree--Fock-plus-BCS (HF+BCS) calculations
with a seniority force using an intrinsic parity-breaking code
developed some years ago \cite{BQS}. The SkM$^{*}$ \cite{Bartel82}
effective interaction is used. The strengths of the seniority force in the $T_z=1$ and $T_z=-1$ channels only
have been fixed \cite{Bonneau2007} for the whole region of
well-deformed nuclei, and a given size of the
valence space in which the BCS equations
are solved, so that neutron and proton odd-even binding-energy
differences (for the ground states) are reasonably well
reproduced. One must keep in mind that the particle-number
non-conserving HF+BCS approach serves merely
here to define relevant one-body auxiliary
potentials $\hat{U}$.

In a second step the strengths of the
$\delta$ residual interaction in the $T_z=1$ and $T_z=-1$ channels have been fixed for a given
nucleus by relying on the phenomenological quality of our HF+BCS
calculations (as hinted, e.g., from by the reasonable results of
Ref.~\cite{BQS} obtained for fission-barrier heights). More
specifically, we have chosen these strengths so as to reproduce the
Fermi-surface diffuseness defined as
$\mathrm{tr}(\hat{\rho}^{1/2}(1-\hat{\rho})^{1/2})$, where $\hat{\rho}$
is the one-body reduced density matrix of the correlated wave function,
obtained in HF+BCS calculations for both charge states. Since this fit
depends, of
course, on the choice of a deformation for the solution, we have
arbitrarily applied this condition at one point in the deformation
space which is of interest for the problem under scrutiny. In our
case, we have chosen to make this fit at the second
fission barrier of
$^{240}$Pu. The HTDA calculations have
then been performed within a space defined:
\begin{itemize}
\item by taking all single-particle states lying in a band of $\pm 6$ MeV
around the proton and neutron Fermi energies in $|\Phi_0\rangle$
(defined as half the differences between the
last occupied and the first unoccupied single-particle
states)
\item by taking all 1-pair transfer states available from the above defined single-particle valence space.
\end{itemize}

Finally, we have carried out a projection after
variation of the obtained HTDA ground state and calculated the
corresponding energy as described above.

%
%

\section{Results and discussion}

The main purpose of this paper is to study the
impact of parity restoration on the outer fission barrier of $^{240}$Pu when starting from a
$K^{\pi}=0^+$ state. This state is relevant for the induced fission of the $1/2^+$
ground state of $^{239}$Pu by an $s$-wave neutron
or the ground-state spontaneous-fission decay of
the $^{240}$Pu nucleus.

To do so we have studied the relevant
characteristics of the two-dimensional deformation energy surface for
axial quadrupole and octupole deformations from before the fission
isomeric state up to after the outer saddle point.

\begin{figure}[h]
\includegraphics[width=10.0cm]{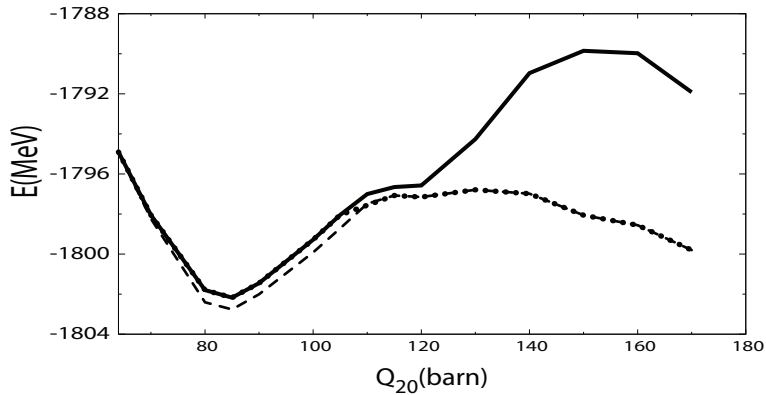}
\caption{Deformation energy curves from before the fission isomeric
  state and up to after the second barrier region of the $^{240}$Pu
  nucleus. Three curves are plotted: for symmetrical HTDA solutions
  (solid curve), for asymmetrical HTDA solutions unprojected (dotted
  curve), for projection on positive parity states (dashed
  curve). \label{N1}}
\end{figure}

The results are summarized in Fig.~\ref{N1}. Throughout this paper, the specific definition of the multipole moments $Q_{20}, Q_{30}$ and $Q_{40}$ is the one given
e.g. in Ref. \cite{Bonneau2005}. As very well known (see
in particular Ref. \cite{BQS} where HF+BCS calculations have been
performed for this nucleus making use of the same Skyrme interaction),
allowing for a breaking of the left-right reflection symmetry
considerably lowers the fission barrier down to values
consistent with their experimental values. Consistently, we find in
our HTDA calculations a value of 5.5 MeV for the second
fission-barrier height to be compared with the 5.1 MeV experimental
value~\cite{Bjornholm1980}.

In that respect one should mention that the conclusion drawn from this
seemingly very good success should be somewhat watered down, a priori,
because these calculations do not take into account some corrections
which would lower the fission barrier height (as the rotational energy
correction, since we are merely calculating here a fission barrier for
intrinsic states) and some others which would raise it. In the latter
case, we mention, beyond the corrections related to the present work,
those found in Refs.~\cite{Anguiano2001,LeBloas2011} and
systematically explained in Ref.~\cite{LeBloas2011} which are stemming
from the approximate treatment of the Coulomb exchange terms.

Let us first remark that upon projecting parity breaking intrinsic
solutions on positive (negative resp.) parity eigenstates, one always
obtains projected energies equal to or lower (resp. higher) than the
energies of the intrinsic
  solutions. The relative ordering of
the two projected energies for the same intrinsic parity breaking
state may be hinted, if not demonstrated, from the following
consideration which is a quite general Quantum Mechanical result (see,
e.g., \cite{COHI1966}). If one were to mix, in a GCM fashion for
instance, positive and negative parity states in the vicinity of the
intrinsic equilibrium solution at a given value of $Q_{20}$, one would
likely generate a zero-node solution in the $Q_{30}$
direction for the lowest positive-parity state and a one-node
solution for the lowest negative-parity state. One knows, e.g., from the discussion of Ref.~\cite{COHI1966}, that the
former would correspond to the ground-state(in such a non-degenerate case). The fact
thus, that positive-parity solutions (not mixed by the GCM ansatz) are
lower in energy than the corresponding negative-parity solutions may
be deemed as consistent with the above. Note also, en passant, that we
have not made the effort to find out energies of negative-parity
states in the vicinity of the reflection-symmetrical case, through
cumbersome appropriate treatments of this limiting case, since we are
merely interested here in the behavior of positive-parity
solutions.

From the above discussion, one can expect four different
  behaviors of the positive-parity energy curve $E^+(Q_{30},
  Q_{20}^{\mathrm{fixed}})$.

(i) One has an equilibrium intrinsic solution which is symmetrical
($Q_{30}=0$) with a large stiffness in the axial octupole mode so
  that the projection is not able to create a
positive-parity minimum for a
  finite value of $Q_{30}$. Clearly, such a projection reduces the stiffness with
respect to the axial octupole mode, possibly creating a dynamical
instability.

\begin{figure}[h]
\includegraphics[width=10.0cm]{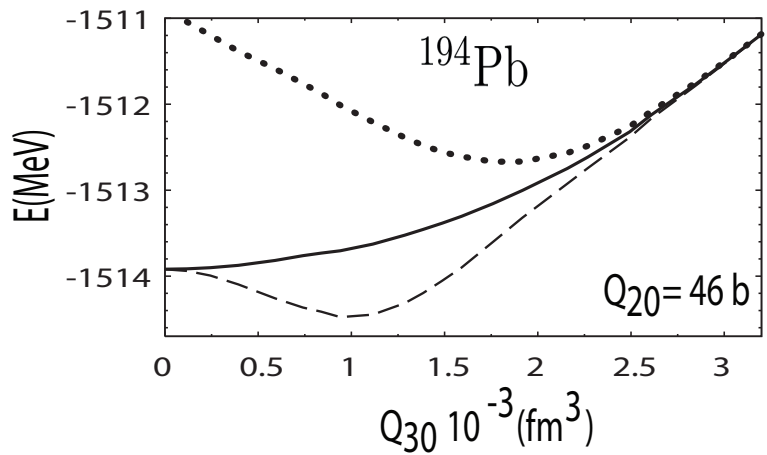}
\caption{Octupole deformation properties of $^{194}$Pb at the SD state
  ($Q_{20} = 46$ b). Three curves are plotted: for asymmetrical
  unprojected HTDA solutions (solid curve), for projection on negative
  parity states (dotted curve), for projection on positive parity
  states (dashed curve). \label{N2}}
\end{figure}

(ii) If, in contrast to (i), the symmetrical equilibrium solution has a small enough
stiffness, the parity projection can create a pocket for a
non-vanishing value of the octupole moment. A static instability away from symmetry, which was not
available with the intrinsic solutions, appears. This
situation has been already encountered in projected HF+BCS
calculations for superdeformed solutions in the Mercury-Lead
region~\cite{Bonche1991}. It has also been found in our
calculations of the super deformed intrinsic state of Pb$^{194}$ as
shown in Fig.~\ref{N2}.

(iii) This case is concerned with situations where the
intrinsic equilibrium solution is not symmetrical but corresponds to a not too
large absolute value of $|Q_{30}|$ (where the concept of large
will be defined in the following case (iv)
below). The positive-parity projection emphasizes
the instability already present at the intrinsic level, yielding
possibly corresponding equilibrium $| Q_{30} |$ values which may
differ significantly from what has been obtained in the intrinsic
case.

(iv) Beyond some critical value of $| Q_{30} |$, the overlap
and hamiltonian kernels between an intrinsic wave function and
its parity transform become negligible, so that in the
calculations of the projected energies according to Eq.~(\ref{Ep}) one obtains
\begin{equation}
E^{(p)} \approx E_{\rm intr} \:,
\end{equation}
where $E_{\rm intr}$ stands for the energy of the intrinsic
  solution $\ket{\Psi}$, namely $E_{\rm intr} = \elmx{\Psi}{\hat
    H}{\Psi}$.

\begin{figure}[h]
\includegraphics[width=10.0cm]{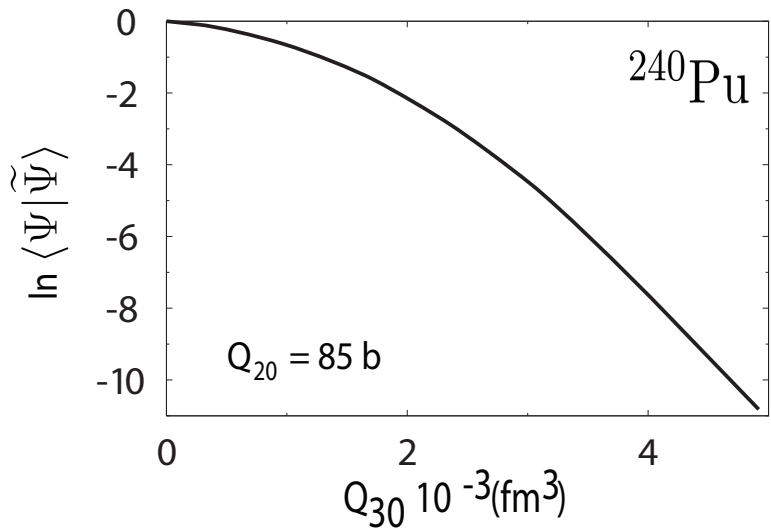}
\caption{Logarithm of the overlap $\overlap{\Psi}{\widetilde{\Psi}}$
  as a function of the octupole deformation at the elongation defined by $Q_{20}=
  85$~b. \label{N3}}
\end{figure}

\begin{figure}[h]
\includegraphics[width=14.0cm]{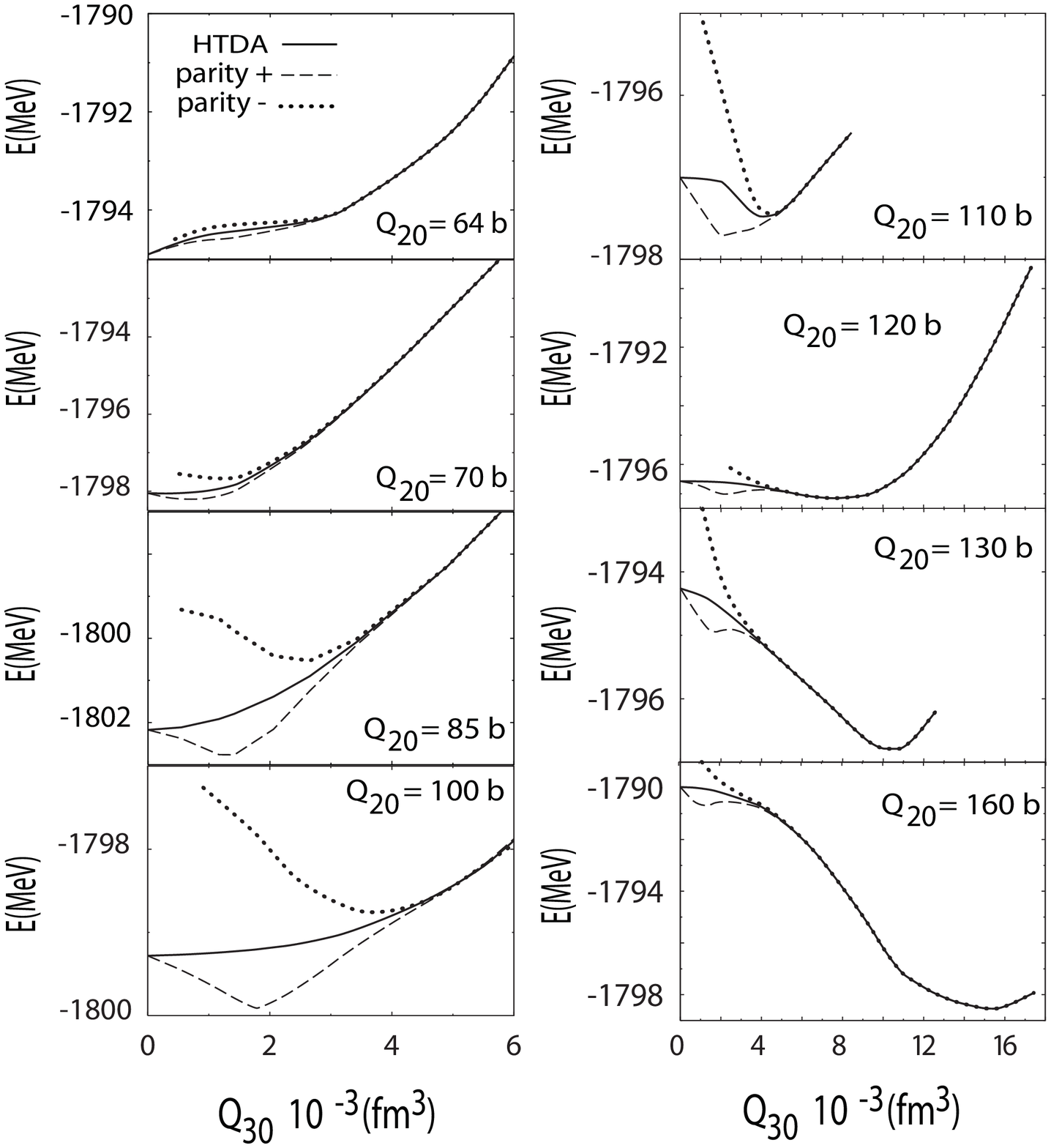}
\caption{Deformation energy curves for some values of the elongation (defined by $Q_{20}$) around
  the isomeric state and the second barrier region of the $^{240}$Pu
  nucleus. Three curves are plotted: for asymmetrical unprojected HTDA
  solutions (solid curve), for the projection on the negative states
  (dotted curve), for the projection on the positive states (dashed
  curve). \label{N5}}
\end{figure}

This is examplified in Fig.~\ref{N3} where the overlap of a solution
obtained for ${}^{240}$Pu nucleus at $Q_{20} = 85$ barns (b)
with parity-transformed wave function is plotted as a
function of $Q_{30}$. The drop in the overlap to about
0.01 just above $Q_{30} \simeq 3.5$ $\mathrm{b^{3/2}}$ is to be
paralleled with the behavior of the energy curves shown
in Fig.~\ref{N5} where a clear convergence
of the three energies $E_{\rm intr}$, $E^{(+)}$ and $E^{(-)}$ is
observed beyond this value of the axial octupole moment.

\begin{figure}[h]
\includegraphics[width=10.0cm]{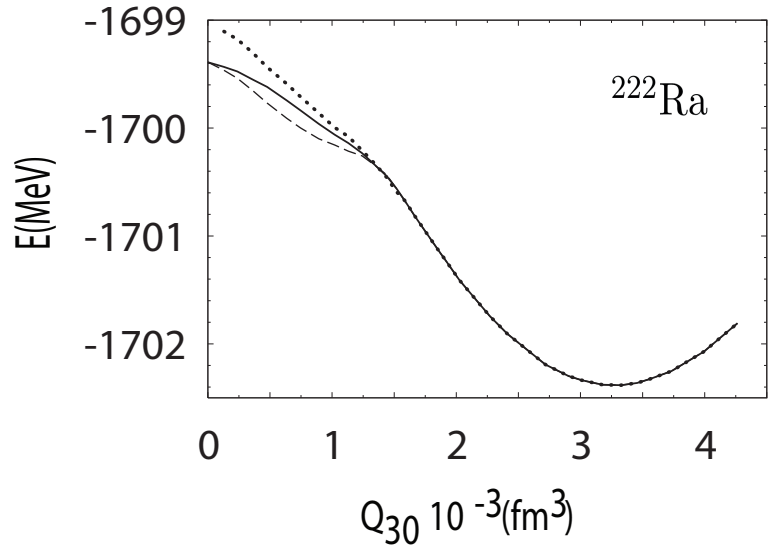}
\caption{Octupole deformation properties of $^{222}$Ra at the ground
  state corresponding to $Q_{20} = 13$ b. Three curves are plotted: for asymmetrical
  unprojected HTDA solutions (solid curve), for projection on negative
  parity states (dotted curve), for projection on positive parity
  states (dashed curve). \label{N4}}
\end{figure}

The consequence of this feature is that when the non-symmetrical
intrinsic equilibrium solution has a value of $| Q_{30} |$ larger than
the above critical value of almost vanishing
overlaps, then the projection is ineffective in producing any change in
the equilibrium solution. One obtains a trivial twofold degeneracy corresponding to solutions which
are connected by parity
  transformation. This situation has also been encountered in
projected Hartree--Fock--Bogolyubov calculations with the Gogny
energy-density functional for ${}^{222}$Ra \cite{Egido1991}. It
has also been found in our calculations for this isotope as
shown on Fig.~\ref{N4}.

\begin{figure}[h]
\includegraphics[width=10.0cm]{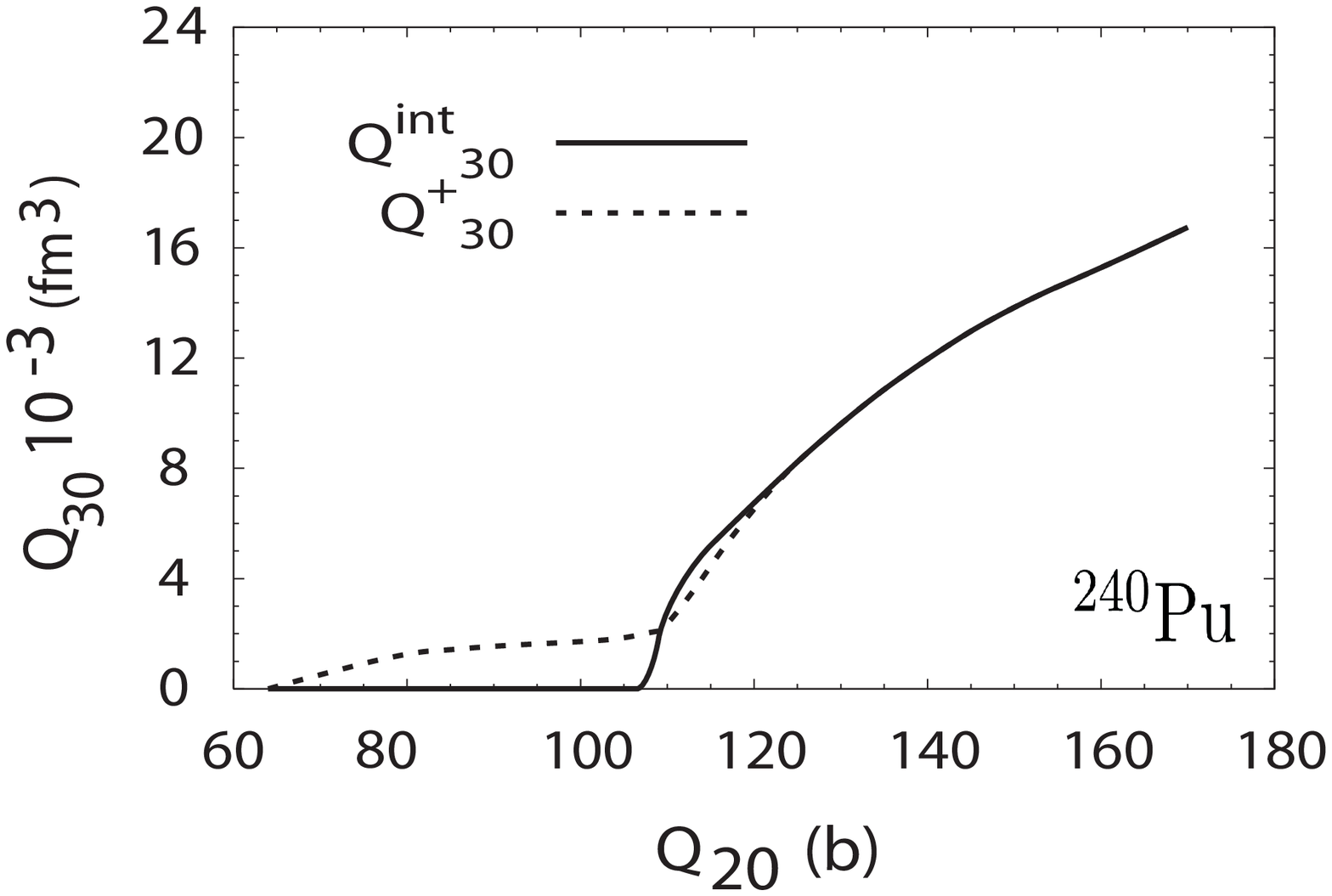}
\caption{The octupole deformation curves of the projection on the
  positive states and asymmetrical unprojected HTDA solutions (denoted
  alternatively by $Q^+_{30}$ and $Q^{\mathrm{int}}_{30}$) as function
  of the quadrupole deformation $Q_{20}$. \label{N6}}
\end{figure}

As can be seen in Fig.~\ref{N5}, we are
encountering the behaviors (i) to (iv) in
  this order in the study of the
minimal energy solutions obtained at a given $Q_{20}$ value, when
increasing it from after the first barrier up to beyond the second
barrier. The transitions between these various patterns
take place at $Q_{20} \simeq 65$, $\simeq 105$ and $\simeq$
125~b. Alternatively we have plotted in Fig.~\ref{N6} the
equilibrium $(Q_{20},Q_{30})$ values for the intrinsic and the
positive parity solutions. Clearly the onset of a stable axial octupole deformation takes
  place in the fission process much earlier for the projected
  solutions than for the intrinsic ones.

This entails two important consequences. First, this leads to the proposed existence of a stable octupole
deformation for the fission isomeric state corresponding to $Q_{20} = 85$ b, $Q_{40} = 10.5$ b$^2$ and $Q_{30} = 4.9$ b$^{3/2}$. Second, this provides a
systematic enhancement of the outer fission
barrier with respect to the fission isomeric
  state. This enhancement is only
due to the relatively soft character of the octupole deformation
energy curve near the fission isomeric state since at the top of the second barrier the
octupole deformation is too large to yield any projection effect.

It is our contention that these two consequences are quite general in the actinide
region. Therefore, we deem that the outer (positive parity) fission-barrier heights with respect to the fission isomeric
  state are probably systematically underestimated in this
region of nuclides, within unprojected calculations. In the present
calculations, the enhancement of the fission barrier height amounts to about 350 keV for the state of the
compound $^{240}$Pu nucleus.

%
%

\section{Conclusions}

In this paper, we have investigated the outer
fission barrier of $^{240}$Pu within the microscopic HTDA approach and
evaluated the effect of the parity projection on
this barrier. The projection on positive-parity states around the
fission isomeric state has demonstrated the instability of this state
with respect to the $Q_{30}$ mode. This confirms the results of Bonche et
al. \cite{Bonche1986} or Egido and Robledo\cite{Egido1991} obtained in another superdeformation
region (superdeformed states in the Hg-Pb region). As well known
already from P. M$\ddot{\mathrm{o}}$ller and S. G. Nilsson, the
fission valley when rising towards the outer
barrier becomes more and more asymmetrical with respect to the
intrinsic parity. Consequently much before reaching the
outer saddle point, one goes beyond the critical values where,
for a given $Q_{20}$ value, the matrix elements
$\langle\Psi|\hat{O}|\widetilde{\Psi}\rangle$ vanish (where $\hat{O}$
stands for the Hamiltonian or the identity operator). In this case it
is clear that projected and unprojected energies are similar. As a
consequence of the isomeric state instability, one expects that the
positive-parity fission-barrier height with respect to the fission isomeric state is larger than the
unprojected one. Therefore, one may conclude that our results provide
a hint for a systematic underestimation of the outer fission barrier height in usual unprojected calculations.



\section{Acknowledgments}

Two of the authors (T. V. N. H. and P. Q.) acknowledge the support by the France Vietnam Particle Physics Laboratory (FV-PPL LIA) and by the National Foundation for Science and Technology Development (NAFOSTED) of Vietnam through Grant No. 103.04-2010.02.


\end{document}